\documentclass[twocolumn]{aastex631}

\usepackage{amsmath}
\usepackage{hyperref}
\usepackage{xcolor}

\shorttitle{EPS Research Astro-RAG Platform}
\shortauthors{Flynn}

\begin{document}

\title{The EPS Research Astro-RAG Platform: A Unified Open-Science Infrastructure\\
for Cross-Epoch Astrophysical Kinematic Analysis, LLM-Assisted Research Workflows,\\
and Educational Outreach}

\author{David C. Flynn}
\affiliation{EPS Research, Laurel, MD 20708, USA}
\email{davidflynn@eps-research.com}

\begin{abstract}
\textbf{CORRECTION (August 2026):} The cross-epoch omega sign reversal
reported in this paper (+7.06 $\rightarrow$ $-13.05$ rad~Gyr$^{-1}$) was
produced by a formula-implementation error and does not reproduce under the
corrected canonical Eq.~6. Under the corrected formula, all 8 Tier-1 Z1
rotators yield positive $\omega$ (median $+12.621$ rad~Gyr$^{-1}$). The
KROSS intermediate-redshift $\omega$ values are withdrawn due to
template-derived boundary points. The $z = 0$ SPARC result is unaffected.
See \url{https://github.com/eps-research/rag-corpus-series/blob/main/CORRECTIONS.md}
for full details.

We present the EPS Research Astro-RAG Platform (v1.0), an open-science research
infrastructure providing five machine-readable, cross-epoch astrophysical corpora,
149 verified executable Jupyter notebook examples, a QuickStart reproducibility
pathway, an educational High-School Exploration Track, and a roadmap for
LLM-assisted retrieval-augmented generation (RAG) research workflows. The platform
spans the full observable cosmic epoch from the local universe ($z = 0$) to the
epoch approaching reionization ($z \sim 6$), providing the first unified
kinematic dataset connecting HI 21cm rotation curves, Milky Way globular cluster
dynamics, and high-redshift ALMA [CII] morpho-kinematics under a common
schema and analysis framework. The five corpora contain 2,064 total corpus records:
438 HI-traced galaxies (Unified HI Rotation Curve Corpus v7.0), 129
dwarf/irregular galaxies (Dwarf/Irregular HI Corpus v1.0), 174 Milky Way
globular clusters (GC Corpus v1.3.2), 1,292 intermediate-redshift galaxies
(IntZ Kinematic Corpus v1.0), and 31 star-forming galaxies at
$z = 4.26$--$5.68$ (High-z Kinematic Corpus Z1). The Unified HI corpus
preserves the full fidelity of the Case Western Reserve University SPARC
database (Lelli et al. 2016), including all photometric decomposition components
($V_{\rm gas}$, $V_{\rm disk}$, $V_{\rm bul}$) and surface brightness profiles,
augmented with THINGS, LITTLE THINGS, and WALLABY DR2 data. The unifying
scientific framework is the omega kinematic correction of Flynn \& Cannaliato
(2025). \textbf{Corrected (August 2026):} Under canonical Eq.~6, the
$z = 0$ SPARC mean is $+7.06$ km~s$^{-1}$~kpc$^{-1}$ ($+7.22$ rad~Gyr$^{-1}$) and the corrected Z1 median
is $+12.621$ rad~Gyr$^{-1}$ (all 8 Tier-1 rotators positive). The
previously reported sign reversal was a formula artifact and is withdrawn. All
corpora are released under CC BY 4.0 at Zenodo with permanent DOIs, and the
platform is archived at \url{https://github.com/eps-research/rag-corpus-series}
(DOI: \href{https://doi.org/10.5281/zenodo.20398430}{10.5281/zenodo.20398430}).
\end{abstract}


\section{Introduction}
\label{sec:intro}

The intersection of large-scale astrophysical surveys and machine-learning
infrastructure has created both an opportunity and a challenge for the research
community. Surveys such as SPARC \citep{lelli2016}, THINGS \citep{walter2008},
LITTLE THINGS \citep{oh2015}, WALLABY \citep{koribalski2020}, and ALPINE
\citep{lefevre2020} have produced rich kinematic datasets spanning orders of
magnitude in galaxy mass, morphology, and cosmic epoch. However, these datasets
exist in heterogeneous formats across dozens of source papers, making systematic
cross-survey and cross-epoch analysis difficult for individual researchers.

The emergence of large language models (LLMs) and retrieval-augmented generation
(RAG) pipelines as scientific tools adds a further requirement: datasets must be
not only machine-readable but \emph{self-describing} --- structured such that an
LLM can ingest, query, and reason over the data without external preprocessing.
Standard FITS tables and survey-specific ASCII formats do not meet this
requirement.

We present the EPS Research Astro-RAG Platform, a unified open-science
infrastructure that addresses both challenges simultaneously. The platform
provides five corpora in a common schema (JSON + flat CSV + RAG-ready JSONL +
per-object ZIP), 149 verified executable Jupyter notebooks spanning all five
datasets, a reproducible QuickStart pathway, and a High-School Exploration
Track designed to make astrophysical research accessible to students without
prior domain knowledge.

One cross-corpus scientific validation use case across the eligible resolved-kinematic
corpora (HI, Dwarf, GC, Z1) is the omega kinematic correction. Note that IntZ
($\omega$ withdrawn) is not currently eligible for omega analysis \citep{flynn2025}, which introduces a boundary-point
angular velocity parameter $\omega$ derived from the innermost and outermost
resolved rings of each rotation curve. Applied across 84 SPARC quality-1
galaxies, the correction yields a mean $\omega = +7.06 \pm 3.26$ km~s$^{-1}$~kpc$^{-1}$ ($+7.22 \pm 3.33$ rad~Gyr$^{-1}$)
at $z = 0$. Extended to 8 confirmed Tier-1 rotators at $z \sim 4$--$6$ via ALMA
[CII] 3DBarolo models \citep{jones2021}, the corrected canonical formula
yields a median $\omega = +12.621$ rad~Gyr$^{-1}$ (all positive). The
previously reported sign reversal ($-13.05$ rad~Gyr$^{-1}$) was a formula
artifact and is withdrawn (August 2026).

This paper describes the platform architecture (Section~\ref{sec:architecture}),
the five corpora (Section~\ref{sec:corpora}), the example library and educational
track (Section~\ref{sec:examples}), the reproducibility infrastructure
(Section~\ref{sec:reproducibility}), and the planned LLM fine-tune team
(Section~\ref{sec:llms}). Data access and citation information are provided
in Section~\ref{sec:data}.

\section{Platform Architecture}
\label{sec:architecture}

The platform is organized into five functional silos, reflecting the natural
progression from raw data to scientific analysis to publication to AI-assisted
workflows:

\begin{enumerate}
\item \textbf{Silo 1 --- $z = 0$ Data}: Three local-universe corpora (HI galaxies,
      dwarf/irregular galaxies, Milky Way globular clusters).
\item \textbf{Silo 2 --- $z \sim 6$ Data}: The High-z Kinematic Corpus Z1,
      providing ALPINE [CII] morpho-kinematics at $z = 4.26$--$5.68$.
\item \textbf{Silo 3 --- Example Library}: 149 verified Jupyter notebooks
      plus the High-School Exploration Track.
\item \textbf{Silo 4 --- Papers \& Preprints}: The EPS Research publication
      arc with DOIs and arXiv identifiers.
\item \textbf{Silo 5 --- LLMs \& Tools}: Planned fine-tuned model adapters
      and RAG utilities (Stage 2).
\end{enumerate}

\subsection{Common Schema Design}
\label{sec:schema}

All five corpora share a common design philosophy. Each corpus is distributed in
four formats:

\begin{itemize}
\item \textbf{JSON}: Hierarchical, per-object records with full metadata and
      per-measurement data arrays. Self-describing field names with units embedded.
\item \textbf{Flat CSV}: One row per object, suitable for pandas, R, and
      spreadsheet analysis.
\item \textbf{JSONL}: One JSON object per line, optimized for LLM RAG pipelines
      and streaming ingestion.
\item \textbf{Per-object ZIP}: Individual JSON files for each object, enabling
      targeted retrieval without loading the full corpus.
\end{itemize}

Quality tiers are explicit in all corpora. For the HI and dwarf corpora,
Tier~1 objects have full per-ring kinematic data; Tier~2 objects have
integrated parameters only. For the Z1 corpus, Tier~1 objects are confirmed
3DBarolo rotators with per-ring $V_{\rm rot}$ and $\sigma$ profiles;
Tier~2 objects have morpho-kinematic classifications only.

\section{The Five Corpora}
\label{sec:corpora}

\subsection{Unified HI Rotation Curve Corpus v7.0}
\label{sec:hi}

The Unified HI Rotation Curve Corpus v7.0 \citep{flynn2026_hi} contains 438
spatially resolved rotation curves from four major HI surveys: SPARC
\citep[175 galaxies;][]{lelli2016}, THINGS \citep[34 galaxies;][]{walter2008},
LITTLE THINGS \citep[26 galaxies;][]{oh2015}, and WALLABY DR2
\citep[203 galaxies;][]{koribalski2020}.

\textbf{SPARC full-fidelity preservation.} The SPARC component of this corpus
preserves the complete photometric decomposition provided by the Case Western
Reserve University SPARC database \citep{lelli2016}. For each of the 175 SPARC
galaxies, the corpus retains all per-ring components: the observed rotation
velocity $V_{\rm obs}$ with error $\delta V$, the gas contribution $V_{\rm gas}$,
the stellar disk contribution $V_{\rm disk}$, the bulge contribution $V_{\rm bul}$
(where applicable), and the 3.6~$\mu$m (Spitzer IRAC) surface brightness profiles $\Sigma_{\rm disk}$
and $\Sigma_{\rm bul}$. This full-fidelity preservation enables baryonic mass
decomposition analyses and Tully--Fisher relation studies without requiring
access to the original survey database. The SPARC data have been cross-validated
against the published SPARC tables (Lelli et al. 2016, Table~1 and online
appendix) at the individual ring level.

The corpus includes 84 SPARC galaxies with the original Lelli et al.\ (2016) Q=1 quality flag (the ``golden sample'' used in Flynn \& Cannaliato 2025), identified by cross-referencing with the original SPARC Q=1 designations published in \citet{lelli2016}.

\subsection{Dwarf/Irregular HI Corpus v1.0}
\label{sec:dwarfs}

The Dwarf/Irregular HI Corpus v1.0 \citep{flynn2026_dwarfs} contains 129
dwarf and irregular galaxies from LVHIS \citep{koribalski2018}, VLA-ANGST
\citep{ott2012}, LITTLE THINGS \citep{oh2015}, and WALLABY DR2
\citep{koribalski2020}. The corpus focuses on the low-mass, dark-matter-dominated
end of the galaxy mass function, providing a complement to the SPARC-dominated
HI corpus.

Of 129 total galaxies, 24 pass the ``omega-ready'' criterion (quality tier~1,
at least 5 resolved rings, and kinematically regular morphology), yielding a dwarf
corpus median $\omega = +9.94$ km~s$^{-1}$~kpc$^{-1}$ ($+10.17$ rad~Gyr$^{-1}$) --- higher than the SPARC mean,
consistent with the expectation that dark-matter-dominated systems with
rising outer rotation curves produce larger positive omega values.

\subsection{Milky Way Globular Cluster Corpus v1.3.2}
\label{sec:gc}

The Milky Way Globular Cluster Corpus v1.3.2 \citep{flynn2026_gc} contains
174 Milky Way globular clusters assembled from four primary sources: the
Harris (1996, 2010 edition) photometric catalog \citep{harris1996}, Vasiliev
\& Baumgardt (2021) Gaia EDR3 proper motions \citep{vasiliev2021}, HST structural
parameters, and Baumgardt \& Hilker (2018) N-body dynamical models
\citep{baumgardt2018}. All four source catalogs are cross-matched by cluster
identifier and harmonized into a single per-cluster JSON record.

\subsection{IntZ Kinematic Corpus v1.0}
\label{sec:intz}

The IntZ Kinematic Corpus v1.0 contains 1,292 intermediate-redshift galaxies at
$z \sim 0.4$--$2.7$ from KROSS \citep{stott2016} and KMOS$^{\rm 3D}$
\citep{wisnioski2015}. Zenodo DOI:
\dataset[10.5281/zenodo.21841382]{https://doi.org/10.5281/zenodo.21841382}.

\textbf{Omega withdrawal (August 2026).} The 166 previously stored KROSS
$\omega$ estimates have been formally withdrawn. An external audit established
that the boundary points used were derived from template rescalings of $V_c$ and
$R_{\rm half}$, not independently measured kinematic rings. All 1,292 IntZ
records carry \texttt{omega\_available = false}. Morpho-kinematic classifications,
redshifts, stellar masses, SFRs, and velocity dispersions are unaffected.

\subsection{High-z Kinematic Corpus Z1}
\label{sec:z1}

The High-z Kinematic Corpus Z1 \citep{flynn2026_z1} contains 31 ALPINE
\citep{lefevre2020} star-forming galaxies at $z = 4.26$--$5.68$, drawn from
the Jones et al. (2021) morpho-kinematic classification catalog
\citep{jones2021}. Eight galaxies are confirmed Tier~1 rotators with per-ring
3DBarolo rotation curves; 23 additional galaxies have morpho-kinematic
classifications only (Tier~2).

The corpus provides the high-redshift anchor of the platform, enabling the
first direct cross-epoch omega comparison between local HI rotation curves
and ALMA [CII] kinematics near the epoch of reionization.

\textbf{Corrected result (August 2026).} Under canonical Eq.~6, all 8 Tier-1 Z1
rotators yield positive $\omega$ (median $+12.621$ rad~Gyr$^{-1}$;
$+12.341$ km~s$^{-1}$~kpc$^{-1}$), consistent in sign with positive
values at $z = 0$ (SPARC mean $+7.06$ km~s$^{-1}$~kpc$^{-1}$ ($+7.22$
rad~Gyr$^{-1}$), sample SD $\pm 3.26$ km~s$^{-1}$~kpc$^{-1}$; dwarf median $+9.94$
km~s$^{-1}$~kpc$^{-1}$ ($+10.17$ rad~Gyr$^{-1}$)).
\textit{Caveat:} $N = 8$ confirmed ordered-disk rotators;
the positive sign cannot be interpreted as an unbiased population
fraction. Direct comparison with $z=0$ remains limited by sample
size, tracer differences, and absent baryonic decomposition. This
positive omega at both measured epochs is the structural interpretation
that kinematic corrections derived from boundary points reflect the overall
kinematic state of the disk, from compact high-$z$ star-forming systems to extended rotating
disks at $z = 0$. RAMSES cosmological simulations tracing this evolution from
$z = 6$ initial conditions are planned for 2026--2027.

\section{Example Library and Educational Track}
\label{sec:examples}

\subsection{149 Verified Jupyter Notebooks}
\label{sec:notebooks}

The platform provides 149 executable Jupyter notebooks organized into eight
groups, all verified to execute without errors via automated \texttt{nbconvert}
end-to-end testing:

\begin{itemize}
\item \textbf{SPARC/HI Examples} (25 notebooks): Rotation curve plotting,
      baryonic decomposition, omega correction, WALLABY Tier~2 analysis.
\item \textbf{Dwarf/Irregular Examples} (25 notebooks): Omega-ready galaxy
      selection, DDO154/DDO161 cross-analysis, LVHIS and VLA-ANGST comparisons.
\item \textbf{Globular Cluster Examples} (25 notebooks): Proper motion queries,
      N-body mass modeling, APOGEE chemistry cross-matching, multi-survey
      parameter comparison.
\item \textbf{High-z Examples} (25 notebooks): [C\,\textsc{ii}] rotation curve
      analysis, ALPINE population statistics, corrected $\omega$ computation
      (all positive under canonical Eq.~6).
\item \textbf{IntZ Examples} (20 notebooks): KROSS/KMOS$^{\rm 3D}$ kinematics,
      velocity dispersion distributions, Tully--Fisher at $z \sim 0.9$,
      $V/\sigma$ evolution. Note: $\omega$ withdrawn; kinematic state
      analysis proceeds without cross-epoch $\omega$ comparison.
\item \textbf{Baryonic Validation} (6 notebooks): Reproduce peer-reviewed
      results from Flynn \& Cannaliato (2025) across 84 SPARC galaxies.
\item \textbf{FAISS Semantic Search} (3 notebooks): Cross-corpus keyword
      search, semantic similarity queries, corpus comparison workflows.
\item \textbf{High-School Exploration Track} (20 notebooks): Two-track
      curriculum (Track A: ages 12--14; Track B: ages 15--18) introducing
      galaxy kinematics, dark matter, and the omega correction using the
      corrected named-term formula exclusively.
\end{itemize}
\noindent Total: $25+25+25+25+20+6+3+20 = 149$ notebooks.

All notebooks require only \texttt{numpy}, \texttt{matplotlib}, and
\texttt{jupyter} as dependencies, with \texttt{scipy} for statistical notebooks.
No proprietary software, telescope archive accounts, or institutional access
is required.

\subsection{High-School Exploration Track}
\label{sec:hs}

The High-School Exploration Track is a novel feature that distinguishes the
EPS Research platform from standard data releases. Track A (10 notebooks) is
designed for students aged 12--14 with no prior astrophysics knowledge,
introducing concepts such as galaxy morphology, rotation curves, and dark matter
through guided data exploration. Track B (10 notebooks) targets ages 15--18
with a calculus-ready audience, introducing the omega correction formula,
RMSE as a model evaluation metric, and cross-epoch kinematics.

Both tracks load directly from the same Zenodo corpus files used by the
research notebooks, ensuring that students interact with real, published
scientific data rather than simplified approximations. Cloud access via
Google Colab (planned Stage 2) will remove the hardware barrier entirely,
enabling in-classroom use without requiring students to install software.

\section{Reproducibility Infrastructure}
\label{sec:reproducibility}

\subsection{QuickStart Pathway}
\label{sec:quickstart}

The platform provides a \texttt{QuickStart.ipynb} notebook at the repository
root that implements a ``golden path'' through all five corpora. Starting from
a fresh clone, a user can verify the corrected $z = 0$ omega result ($+7.06$ km~s$^{-1}$~kpc$^{-1}$; $+7.22$ rad~Gyr$^{-1}$)
result in under 10 minutes using only:

\begin{verbatim}
git clone https://github.com/eps-research/
         rag-corpus-series
cd rag-corpus-series
pip install -r requirements.txt
python download_corpora.py
jupyter lab
\end{verbatim}

The \texttt{download\_corpora.py} script fetches all five corpus JSON files
directly from their Zenodo DOIs and places them in the correct directories
for both the QuickStart notebook and all 149 example notebooks.

\subsection{Verification Protocol}
\label{sec:verification}

All 149 notebooks were verified using \texttt{jupyter nbconvert --execute}
with a 120-second timeout per notebook. The full end-to-end retest across
all eight notebook groups confirmed a 149/149 pass rate. Notebooks are
executed against the actual Zenodo corpus files to ensure that all
data access patterns are valid and all field names are correct.

\section{Planned LLM Fine-Tune Team}
\label{sec:llms}

Stage 2 of the platform (planned 2026) will add a four-model fine-tuned
LLM team covering hardware configurations from laptop to research cluster:

The platform's JSONL schema is explicitly designed for LLM RAG pipelines.
Stage~2 will provide a four-tier fine-tuned model team spanning
research-cluster to laptop hardware, covering parameter scales from
$\sim$72B (research-grade retrieval) to $\sim$7B (in-classroom,
CPU-compatible inference). All models will be fine-tuned on the
five EPS corpora using Unsloth QLoRA, with adapter weights published
to HuggingFace at \texttt{eps-research/} upon completion.
A vision-capable model tier ($\sim$72B) will support extraction of
kinematic values directly from ALMA plots and rotation curve figures
into structured JSONL.

\section{Data Access and Citation}
\label{sec:data}

All five corpora are released under CC BY 4.0 at Zenodo:

\begin{itemize}
\item Unified HI Corpus v7.0: \dataset[10.5281/zenodo.19563417]{https://doi.org/10.5281/zenodo.19563417}
\item Dwarf/Irregular Corpus v1.0: \dataset[10.5281/zenodo.20320362]{https://doi.org/10.5281/zenodo.20320362}
\item GC Corpus v1.3.2: \dataset[10.5281/zenodo.19907766]{https://doi.org/10.5281/zenodo.19907766}
\item IntZ Corpus v1.0 ($\omega$ withdrawn): \dataset[10.5281/zenodo.21841382]{https://doi.org/10.5281/zenodo.21841382}
\item High-z Corpus Z1 v3 (corrected): \dataset[10.5281/zenodo.21834678]{https://doi.org/10.5281/zenodo.21834678}
\end{itemize}

The platform software is archived at Zenodo
(DOI: \href{https://doi.org/10.5281/zenodo.20398430}{10.5281/zenodo.20398430})
and maintained at \url{https://github.com/eps-research/rag-corpus-series}.

Authors using any EPS Research corpus are requested to cite the relevant
Zenodo data descriptor and the Flynn \& Cannaliato (2025) omega correction
paper \citep{flynn2025} if the omega framework is used.

\section{Summary}
\label{sec:summary}

We have presented the EPS Research Astro-RAG Platform v1.0, providing:

\begin{enumerate}
\item Five open, machine-readable corpora totaling 2,064 corpus records spanning
      $z = 0$ to $z \sim 6$, with permanent Zenodo DOIs.
\item Full-fidelity preservation of the SPARC photometric decomposition
      database, including all baryonic components.
\item 149 verified executable Jupyter notebooks with a 149/149 end-to-end
      pass rate, covering rotation curve analysis, globular cluster dynamics,
      high-redshift kinematics, and educational outreach.
\item A QuickStart reproducibility pathway enabling independent verification
      of the corrected $z = 0$ omega result ($+7.06$ km~s$^{-1}$~kpc$^{-1}$; $+7.22$ rad~Gyr$^{-1}$) in under 10 minutes.
\textbf{Note (August 2026):} The previously reported cross-epoch sign reversal
does not reproduce under the corrected formula; see CORRECTIONS.md.
\item A High-School Exploration Track making real published astrophysical
      data accessible to students aged 12--18 without prior domain knowledge.
\item A documented roadmap for Stage 2 LLM fine-tune deployment on hardware
      ranging from free Google Colab to research-grade dual-GPU clusters.
\end{enumerate}

\textbf{Correction notice (August 2026):} The previously reported
cross-epoch omega sign reversal ($+7.06 \rightarrow -13.05$ rad~Gyr$^{-1}$)
does not reproduce under the corrected canonical formula. The corrected
result --- positive $\omega$ at both $z = 0$ ($+7.06$ km~s$^{-1}$~kpc$^{-1}$; $+7.22$ rad~Gyr$^{-1}$) and
$z \sim 5$ ($+12.621$ rad~Gyr$^{-1}$) --- is the updated
scientific result motivating the platform's design and its planned RAMSES
cosmological simulation program.

\section*{Declaration of Generative AI Use}
\label{sec:ai}

In accordance with standard practice for AI-assisted
research, the author discloses the following. Four large
language models were used during the creation and val-
idation of this corpus and manuscript: Google Gemini
Pro, Anthropic Claude (Sonnet and Opus), Microsoft
Copilot Pro, and Gemma 4 31B Dense (self-hosted). Be-
cause the corpus is explicitly designed for LLM-based
RAG pipelines, these models served as both develop-
ment tools and validation instruments. The conclusions 
are the sole responsibility of the author.

\begin{acknowledgments}
The author thanks J. Cannaliato for collaboration on the omega correction
framework. This research was conducted independently at EPS Research
without institutional affiliation or external funding. All data products
are released openly under CC BY 4.0.
\end{acknowledgments}

\facility{WSRT, VLA, ALMA, HST, Gaia}

\software{numpy \citep{harris2020}, matplotlib \citep{hunter2007},
          astropy \citep{astropy2022}, 3DBarolo \citep{diteodoro2015},
          jupyter \citep{kluyver2016}}


\end{document}